\documentclass[conference]{IEEEtran}
\IEEEoverridecommandlockouts
\usepackage{cite}
\usepackage{amsmath,amssymb,amsfonts}
\usepackage{algorithmic}
\usepackage{array}
\usepackage{graphicx}
\usepackage{textcomp}
\usepackage{xcolor}
\usepackage{caption}
\usepackage{multirow}
\usepackage[table]{xcolor}
\usepackage{colortbl}
\usepackage{subcaption}
\usepackage{xspace}
\usepackage{comment}
\usepackage{hyperref}
\usepackage{fontawesome5}
\def\BibTeX{{\rm B\kern-.05em{\sc i\kern-.025em b}\kern-.08em
    T\kern-.1667em\lower.7ex\hbox{E}\kern-.125emX}}
\def\eg{\emph{e.g.,}\xspace}

\def\ie{\emph{i.e.,}\xspace}
\def\etal{\emph{et al.}\xspace}
\def\vs{\emph{vs.}\xspace}

\begin{document}

\title{DUAP: Dual-task Universal Adversarial Perturbations Against Voice Control Systems
\vspace{-4mm}
}

\author{
\IEEEauthorblockN{
Suyang Sun$^{1}$, Weifei Jin$^{1}$, Yuxin Cao$^{2}$, Wei Song$^{3,4}$, Jie Hao$^{1,\href{mailto:haojie@bupt.edu.cn}{\text{\faEnvelope[regular]}}}$
}
\IEEEauthorblockA{
\textit{$^{1}$Beijing University of Posts and Telecommunications, China \quad
$^{2}$National University of Singapore, Singapore} \\
\textit{$^{3}$University of New South Wales, Australia \quad
$^{4}$CSIRO's Data61, Australia} \\
{\small \texttt{\{sunsuyang2023, weifeijin, haojie\}@bupt.edu.cn, yuxincao@u.nus.edu, wei.song1@unsw.edu.au}}
}
}
\maketitle

\begin{abstract}
Modern Voice Control Systems (VCS) rely on the collaboration of Automatic Speech Recognition (ASR) and Speaker Recognition (SR) for secure interaction. However, prior adversarial attacks typically target these tasks in isolation, overlooking the coupled decision pipeline in real-world scenarios. Consequently, single-task attacks often fail to pose a practical threat. To fill this gap, we first utilize gradient analysis to reveal that ASR and SR exhibit no inherent conflicts. Building on this, we propose Dual-task Universal Adversarial Perturbation (DUAP). Specifically, DUAP employs a targeted surrogate objective to effectively disrupt ASR transcription and introduces a Dynamic Normalized Ensemble (DNE) strategy to enhance transferability across diverse SR models. Furthermore, we incorporate psychoacoustic masking to ensure perturbation imperceptibility. Extensive evaluations across five ASR and six SR models demonstrate that DUAP achieves high simultaneous attack success rates and superior imperceptibility, significantly outperforming existing single-task baselines. Our source code is available at~\url{https://github.com/Susuyyyy1/DUAP}.

\end{abstract}


\section{Introduction}
\label{sec:intro}
Voice Control Systems (VCS) have rapidly become ubiquitous in modern daily life, serving as the core interface for a wide range of applications, from smart home devices and autonomous vehicles to mobile personal assistants~\cite{bhanushali2024adversarial}. These systems offer a seamless, hands-free interaction experience by integrating two fundamental components: Automatic Speech Recognition (ASR)~\cite{zhang2024laseradv, jin2025whispering}, which transcribes spoken commands into executable text, and Speaker Recognition (SR)~\cite{yu2023smack, chen2023qfa2sr}, which acts as a biometric gatekeeper to verify user identity and prevent unauthorized access. Despite their widespread adoption and convenience, the security of VCS has emerged as a critical concern~\cite{chi2024adversarial}. More broadly, recent studies have shown that deep neural networks deployed in safety-critical perceptual systems, including audio and video applications, are inherently vulnerable to adversarial examples~\cite{carlini2018audio,jinalmguard,zhong2025synerguard,song2024correction,jin2025boosting,song2025vidtoken}. By injecting carefully crafted perturbations into the audio input, attackers can successfully manipulate the decision logic of these models, rendering ASR systems incapable of recognizing valid commands while simultaneously misleading SR systems into authenticating unauthorized identities. These vulnerabilities raise significant security alarms for safety-critical voice applications, where such breaches could lead to unauthorized access bypass, system malfunction, or critical service unavailability~\cite{chen2023qfa2sr, carlini2018audio, zhang2017dolphinattack}.

Although existing audio adversarial attacks have effectively revealed vulnerabilities in mainstream speech models on tasks like ASR and SR~\cite{yu2023smack, chen2023qfa2sr, fang2024zero}, existing methods suffer from a critical limitation: they predominantly focus on attacking a single task in isolation.
In real-world VCS scenarios, however, these two components operate in a collaborative manner. A voice command is executed only when it satisfies two conditions simultaneously: the content is correctly transcribed by the ASR module, and the user identity is successfully verified by the SR module. Consequently, conventional single-task attacks are insufficient to pose a genuine threat to modern VCS, as the system effectively blocks the intrusion if either the semantic check or the identity check fails.

\begin{figure}[t]
    \centering
    \includegraphics[width=\linewidth]{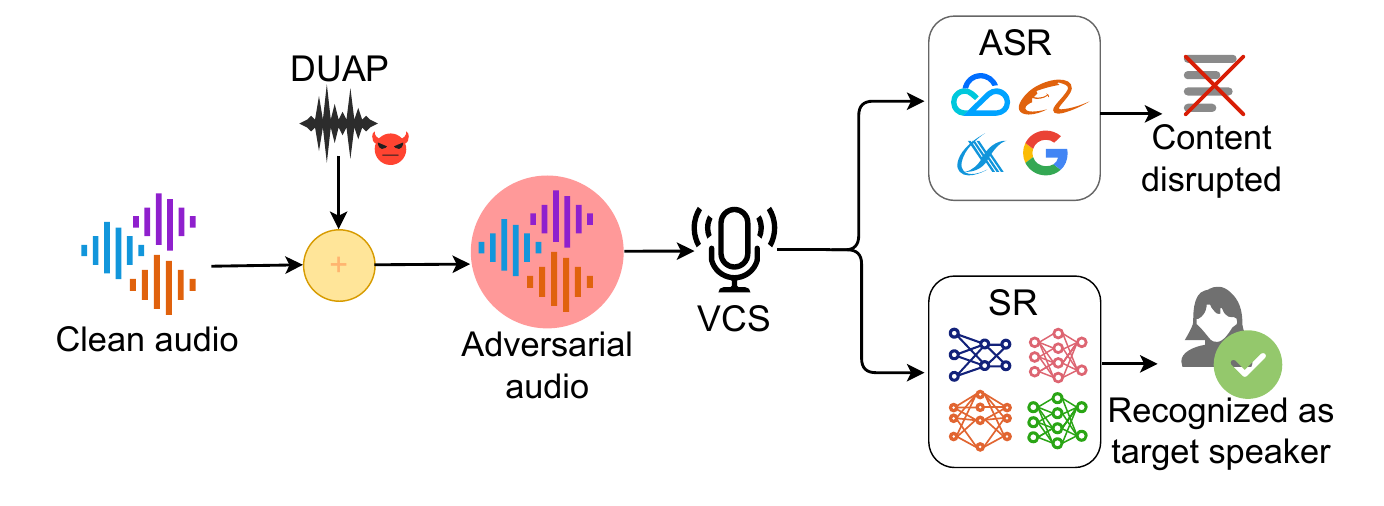}
    \caption{Illustration of our attack scenario. 
    }
    \label{fig:attack_scenario}
\end{figure}

To bridge this gap, we first conduct a comprehensive analysis of the interaction between ASR and SR models using gradient analysis. Our investigation reveals a critical insight: these two tasks exhibit no inherent conflicts, implying that it is feasible to optimize adversarial perturbations that satisfy the objectives of both tasks simultaneously. Building on this finding, we propose \textbf{D}ual-task \textbf{U}niversal \textbf{A}dversarial \textbf{P}erturbation (\textbf{DUAP}), a novel attack framework capable of generating coordinated black-box attacks against the entire VCS pipeline, as illustrated in Fig.~\ref{fig:attack_scenario}. To further enhance the attack's practicality, we introduce a Dynamic Normalized Ensemble (DNE) training strategy to boost the transferability of adversarial examples across different SR models, and we incorporate a psychoacoustic masking constraint to improve the stealthiness of the generated adversarial examples.

To validate the effectiveness of our proposed method, we conduct extensive experiments on diverse datasets across five ASR models and six SR models, including commercial APIs such as Tencent, Alibaba, and iFlytek. The experimental results demonstrate that DUAP achieves high attack success rates on both tasks simultaneously, significantly outperforming single-task baselines such as Zong \etal~\cite{zong2021targeted} and Hanina \etal~\cite{hanina2024universal}. Notably, our method exhibits strong black-box transferability, maintaining near-perfect performance on unseen SR models (\eg HuBERT~\cite{hsu2021hubert}, X-vector~\cite{snyder2018x}) and commercial ASR systems. Furthermore, evaluations using both Signal-to-Noise Ratio (SNR) and Mean Opinion Score (MOS) confirm that DUAP achieves a superior balance between attack strength and audio quality compared to existing ASR-oriented attacks. 
These results highlight that DUAP successfully compromises the dual-task decision pipeline, posing a realistic and severe threat to modern voice control systems.

Our main contributions are summarized as follows:
\begin{itemize}
    \item We propose a dual-task adversarial attack framework targeting the complete VCS pipeline, and we reveal the feasibility of simultaneous attacks through gradient
    analysis.
    \item We propose DUAP to realize this attack, incorporating a DNE strategy to enhance transferability and a psychoacoustic masking constraint to ensure high stealthiness.
    \item We conduct extensive experiments on five ASR and six SR models, demonstrating that DUAP outperforms single-task baselines by achieving robust cross-task generalization and superior audio quality.
\end{itemize}

\section{Related Work}

\subsection{Audio Attacks on Automatic Speech Recognition}
ASR systems focus on decoding phonetic and semantic information from continuous audio signals.
Early attacks primarily targeted specific audio instances. Recent SOTA methods~\cite{wu2023kenku, cheng2024alif, fang2024zero} have advanced this direction by employing refined optimization strategies to penetrate commercial ASR systems with high efficiency or without accessing gradients. While effective, these instance-specific methods incur high computational costs for each input, limiting real-time deployment.
To address this efficiency bottleneck, researchers proposed Universal Adversarial Perturbations (UAPs) that add a single fixed noise to arbitrary inputs. Neekhara \etal~\cite{neekhara2019universal} introduced untargeted UAPs to increase error rates, while Zong \etal~\cite{zong2021targeted} developed targeted UAPs to force specific transcriptions. More recently, AdvDDoS~\cite{ge2023advddos} leveraged local surrogate ensembles to achieve robust untargeted transferability. However, these methods are confined to the ASR domain. Our work adopts a similar surrogate-based strategy but extends the objective to simultaneously compromise the SR system.

\subsection{Audio Attacks on Speaker Recognition}
Unlike ASR, SR aims to extract speaker-dependent characteristics. Depending on the application scenario, SR tasks are typically categorized into three protocols: Speaker Verification (SV), Closed-Set Identification (CSI), and Open-Set Identification (OSI). SV determines whether a specific utterance belongs to a claimed identity. CSI classifies an input as one of the multiple enrolled speakers. OSI extends the closed-set scenario by enabling the system to reject unknown speakers.
Attacks in this domain often manipulate speaker embeddings to cause misclassification. For instance, QFA2SR~\cite{chen2023qfa2sr} and SMACK~\cite{yu2023smack} utilize transfer-based or semantically meaningful perturbations to mislead black-box SR models. Although these attacks demonstrate strong performance, they typically require generating a unique perturbation for each utterance or rely on heavy computation, which hinders practical real-time usage.
To enable broader applicability, Universal attacks have been proposed to generate a single robust perturbation. Xie \etal~\cite{xie2020real} proposed a real-time universal attack capable of impersonating a target identity over-the-air, while Hanina \etal~\cite{hanina2024universal} introduced a stealthy, speaker-independent attack that degrades verification scores. 
Despite their success, these universal methods are exclusively designed for SR and fail to disrupt ASR systems. In contrast, our proposed DUAP bridges this gap by learning a unified perturbation that performs a targeted attack on SR under the CSI setting while simultaneously disrupting ASR transcription.

\section{Methodology}

\subsection{Design Intuition}
Speech signals are inherently multifaceted, encoding a rich spectrum of information including linguistic content, timbre, pitch, rhythm, emotion, among others. Different speech processing tasks typically focus on distinct subsets of these acoustic features. Specifically, ASR models prioritize the extraction of linguistic content (\eg phonemes) while suppressing speaker-dependent variability, whereas SR models focus on capturing identity-related embeddings (\eg timbre) while largely ignoring phonetic variations.
Intuitively, this distinct feature selection suggests that ASR and SR operate on largely non-overlapping feature subspaces. Therefore, it should be theoretically feasible to inject a unified perturbation that manipulates both sets of features simultaneously without creating significant interference between the two objectives.

\begin{figure}[t]
    \centering
    \includegraphics[width=0.68\linewidth]{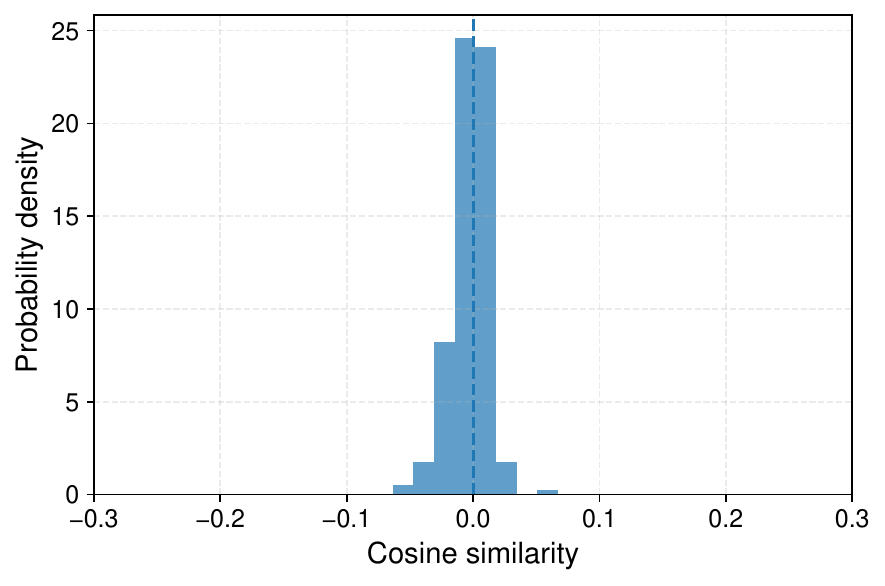}
    \caption{Distribution of cosine similarity between ASR and SR gradients. The concentration around zero indicates that the optimization directions for the two tasks are nearly orthogonal, validating the feasibility of joint optimization.}
    \label{fig:grad_cos_distribution}
    \vspace{-1.0em}
\end{figure}

To rigorously verify this intuition from an optimization perspective, we analyze the geometric relationship between the adversarial gradients of the two tasks. Let $\delta$ denote the universal perturbation. We compute the cosine similarity between the gradients $\nabla_{\delta} \mathcal{L}_{\text{ASR}}$ and $\nabla_{\delta} \mathcal{L}_{\text{SR}}$ with respect to the input perturbation. As illustrated in Fig.~\ref{fig:grad_cos_distribution}, the distribution of cosine similarities is sharply concentrated around zero. This empirical observation confirms that the optimization directions for ASR and SR are mathematically \textbf{orthogonal} in the high-dimensional input space. This orthogonality validates our hypothesis that the two tasks are computationally compatible, allowing DUAP to jointly optimize for both objectives without suffering from gradient conflict.

\subsection{Problem Formulation}

We formulate the dual-task universal adversarial attack as a constrained optimization problem maximizing the joint likelihood of the target events. Let $\mathcal{X}$ denote the distribution of benign speech inputs. Given an ASR model $\theta_{\text{ASR}}$ and an SR model $\theta_{\text{SR}}$, we designate a specific target text transcription $y^*$ and a target speaker identity $s^*$.

Our goal is to find a single universal perturbation $\delta$ that, when added to any input $x \sim \mathcal{X}$, maximizes the probability that the perturbed audio $x_{\text{adv}} = x + \delta$ satisfies both target conditions simultaneously. Assuming the predictions of the ASR and SR models are conditionally independent given the input, the problem can be formulated as maximizing the expected joint probability:
\begin{equation}
\begin{aligned}
\max_{\delta} \quad & \mathbb{E}_{x \sim \mathcal{X}} \Big[ P(y^* \mid x+\delta; \theta_{\text{ASR}}) \cdot P(s^* \mid x+\delta; \theta_{\text{SR}}) \Big] \\
\text{s.t.} \quad & \|\delta\|_{\infty} \le \epsilon,
\end{aligned}
\end{equation}
where $P(\cdot \mid \cdot)$ represents the conditional probability predicted by the respective models, and $\epsilon$ bounds the $\ell_\infty$-norm of the perturbation to ensure imperceptibility.

Directly maximizing this joint probability is numerically unstable. Therefore, in practice, we convert this problem into minimizing the negative log-likelihood loss.

\subsection{Overview of DUAP}
\begin{figure}[t]
    \centering
      \resizebox{0.8\columnwidth}{!}{\includegraphics{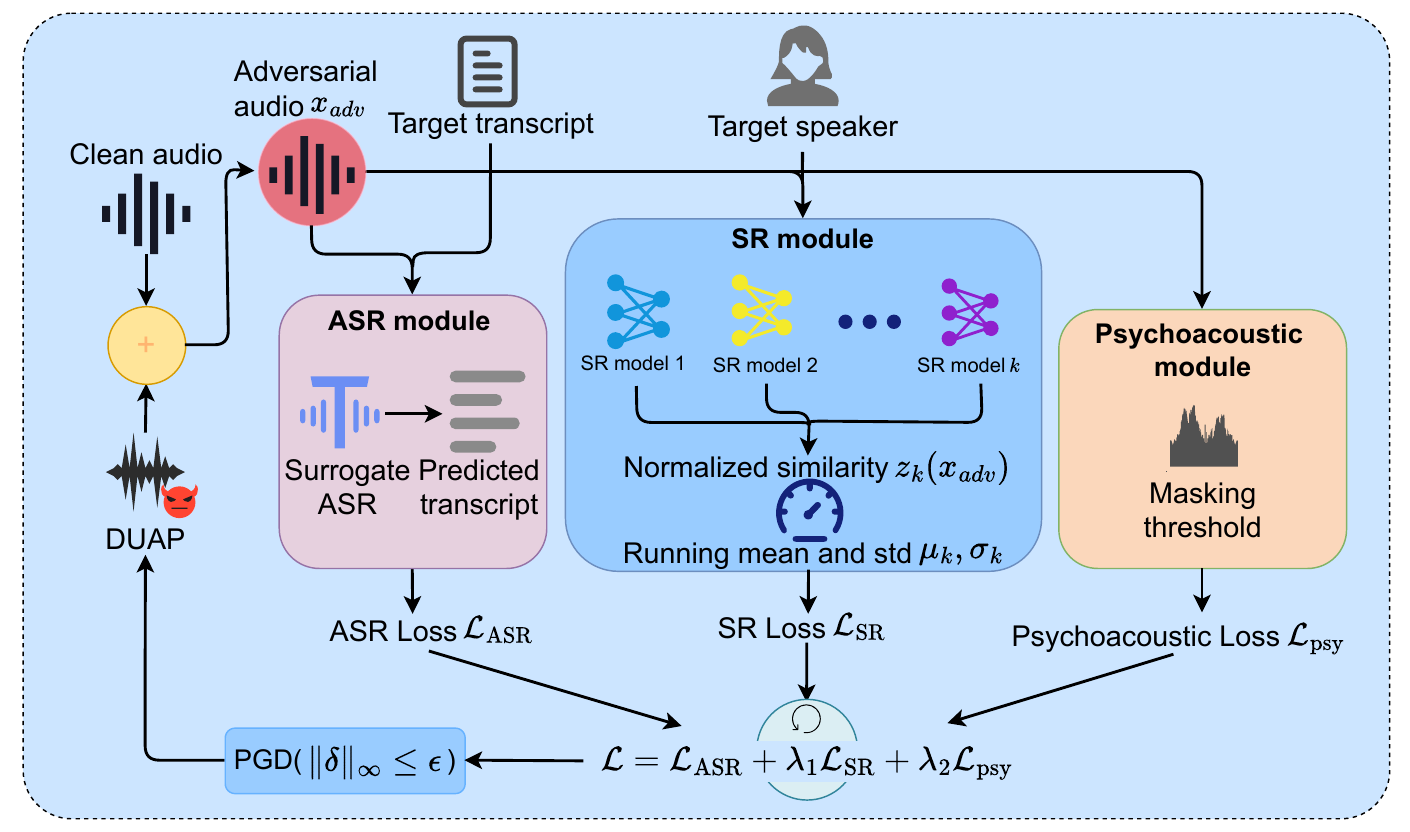}}
    \caption{Overview of the proposed dual-task universal adversarial attack framework. A single perturbation~$\delta$ is optimized with ASR, SR, and psychoacoustic objectives, enabling simultaneous disruption of ASR content and targeted speaker impersonation, while generalizing across speakers, utterances, and unseen models.}
    \label{fig:framework}
    \vspace{-1.0em}
\end{figure}

To solve the formulation above, we propose the DUAP framework, as illustrated in Fig.~\ref{fig:framework}. We optimize the universal perturbation $\delta$ by minimizing a unified objective function composed of three strategic components.

\noindent\textbf{Targeted ASR Optimization.} 
To enhance the transferability of \textit{untargeted} disruption, we employ a \textit{targeted} optimization strategy on local surrogates. This approach stabilizes gradient directions, yielding significantly more robust attacks against black-box APIs compared to naive untargeted maximization.

\noindent\textbf{Dynamic Normalized Ensemble (DNE) for SR.} 
To address optimization instability caused by heterogeneous SR models with varying loss scales, we introduce the DNE strategy. It dynamically balances gradient contributions, ensuring the perturbation consistently drives inputs toward the target identity across diverse embedding spaces.

\noindent\textbf{Psychoacoustic Regularization.} 
Empirically, robust universal ASR attacks require a significantly larger perturbation budget than SR attacks, rendering standard $\ell_\infty$ constraints insufficient for maintaining audio quality. We thus incorporate psychoacoustic masking to hide high-energy perturbations below human auditory thresholds, ensuring imperceptibility.

\subsection{Dual-Task Adversarial Optimization}
\subsubsection{Targeted ASR Optimization}
To achieve the ASR objective, we aim to maximize the log-likelihood of the target transcription $y^*$. For a given perturbed input $x_{\text{adv}}$, the ASR model produces a probability distribution over token sequences. We adopt the standard Connectionist Temporal Classification (CTC) or Cross-Entropy loss (depending on the ASR architecture) to force the model's prediction to align with $y^*$. The loss is defined as:
\begin{equation}
\mathcal{L}_{\mathrm{ASR}} = - \mathbb{E}_{x \sim \mathcal{X}} \left[ \log P(y^* \mid x + \delta; \theta_{\text{ASR}}) \right].
\end{equation}
Minimizing this loss effectively drives the linguistic representation of all inputs toward the semantic space of the target phrase $y^*$, disrupting the original content.

\subsubsection{Dynamic Normalization Ensemble for SR}
For the SR task, our goal is to drive all perturbed speech samples toward a unified target speaker in the embedding space, ensuring that they are consistently classified as the target identity under the CSI setting, where the system is constrained to identify the input as one of the pre-enrolled speakers.

To enhance transferability across varying feature manifolds, we employ an ensemble of $K$ heterogeneous SR models. For each model, we pre-compute the representations for all speakers in the enrollment set. Specifically, let $p_{k,j}$ denote the prototype vector (\ie~the centroid of enrollment embeddings) for the $j$-th speaker extracted by the $k$-th model. 
Let $e_k(x_{\mathrm{adv}})$ be the normalized embedding of the adversarial audio extracted by model $f_k$. The similarity score (logit) between the perturbed audio and the $j$-th speaker is computed as:
\begin{equation}
z_{k,j} = e_k(x_{\mathrm{adv}})^\top p_{k,j}.
\end{equation}
Based on these logits, the raw cross-entropy loss for the target speaker $s^*$ on the $k$-th model is:
\begin{equation}
\mathcal{L}_k^{\mathrm{raw}} = - \log \frac{\exp(z_{k,s^*})}{\sum_j \exp(z_{k,j})}.
\end{equation}

However, directly summing losses from heterogeneous SR models (\eg ECAPA-TDNN~\cite{desplanques2020ecapa} \vs WavLM~\cite{chen2022wavlm}) often leads to optimization instability due to varying loss scales and convergence speeds. To mitigate this \textit{gradient dominance} issue, we introduce a DNE loss. 
Let $i$ denote the current iteration step during the optimization. We maintain the running mean $\mu_k$ and the second moment $\sigma_k$ of the loss for each model, updated via Exponential Moving Average (EMA) with momentum $m$:
\begin{equation}
\left\{ \begin{array}{l}
\begin{aligned}
\mu_k^{(i)} &= m\,\mu_k^{(i-1)} + (1-m)\,\mathcal{L}_k^{\mathrm{raw}(i)}, \\
\sigma_k^{(i)} &= m\,\sigma_k^{(i-1)} + (1-m)\left(\mathcal{L}_k^{\mathrm{raw}(i)}\right)^2.
\end{aligned}
\end{array} \right.
\end{equation}
Using these statistics, we estimate the instantaneous standard deviation as $\mathrm{std}_k^{(i)} = \sqrt{\sigma_k^{(i)} - (\mu_k^{(i)})^2 + \epsilon}$. The loss is then dynamically normalized:
\begin{equation}
\widetilde{\mathcal{L}}_k^{(i)} = \frac{\mathcal{L}_k^{\mathrm{raw}(i)} - \mu_k^{(i)}}{\mathrm{std}_k^{(i)}}.
\end{equation}

To prevent negative gradients when a model's loss drops below its historical mean, we apply a non-negative truncation: $\mathcal{L}_k^{\mathrm{final}} = \max(0, \widetilde{\mathcal{L}}_k^{(i)})$.
Finally, the ensemble SR loss is defined as the average over $K$ models:
\begin{equation}
\mathcal{L}_{\mathrm{SR}} = \frac{1}{K} \sum_{k=1}^{K} \mathcal{L}_k^{\mathrm{final}}.
\end{equation}
By adaptively balancing gradient contributions, DNE ensures that the perturbation generalizes effectively across distinct feature spaces.

\subsubsection{Psychoacoustic Regularization}
To improve stealthiness beyond simple $\ell_\infty$ constraints, we incorporate a psychoacoustic masking constraint. 
The human auditory system has varying sensitivity across different frequency bands, where high-energy components (maskers) can render nearby noise inaudible.~\cite{zwicker2013psychoacoustics}
We compute a frequency-dependent masking threshold $T(t,f)$ for the clean audio $x$ using a psychoacoustic model, where $t$ and $f$ denote the time frame and frequency bin indices, respectively. The perturbation in the Short-Time Fourier Transform domain, denoted as $X_{\delta}(t,f)$, is penalized only when its magnitude exceeds this threshold:
\begin{equation}
\mathcal{L}_{\mathrm{psy}} = \sum_{t,f} \max\big(0,\; |X_{\delta}(t,f)| - T(t,f)\big).
\end{equation}
This loss effectively constrains the perturbation energy to ``hide'' under the perceptible threshold of the original speech, significantly improving audio quality.

\subsubsection{Overall Objective}
We jointly optimize the universal perturbation $\delta$ by minimizing the weighted sum of the ASR, SR, and psychoacoustic losses subject to the $\ell_\infty$-norm constraint. The final optimization problem is formulated as:
\begin{equation}\label{eq:optimization}
\min_{\|\delta\|_{\infty} \le \epsilon} \Big( \mathcal{L}_{\mathrm{ASR}} + \lambda_1 \mathcal{L}_{\mathrm{SR}} + \lambda_2 \mathcal{L}_{\mathrm{psy}} \Big),
\end{equation}
where $\lambda_1$ and $\lambda_2$ are hyperparameters balancing the dual-task trade-offs.

To solve this, we employ Projected Gradient Descent (PGD)~\cite{madry2018towards}. Let $\mathcal{L}(\delta)$ denote the objective function in Eq.~\eqref{eq:optimization}. The perturbation is iteratively updated as:
\begin{equation}
\delta^{(i+1)} = \Pi_{\epsilon} \left( \delta^{(i)} - \alpha \cdot \mathrm{sign}\left(\nabla_{\delta} \mathcal{L}(\delta^{(i)})\right) \right),
\end{equation}
where $\alpha$ is the step size, and $\Pi_{\epsilon}(\cdot)$ projects the perturbation onto the $\ell_\infty$-ball to ensure imperceptibility.

\section{Experiments}

\subsection{Experimental Setup}

\noindent\textbf{Datasets and Models.} 
We optimize the universal perturbation using 500 randomly selected clips from the LibriSpeech~\cite{panayotov2015librispeech} \textit{train-clean} subset.
For ASR, we utilize Whisper-small~\cite{radford2023robust} as the surrogate model. Evaluation is conducted on 150 samples from the \textit{test-clean} subset against DeepSpeech2~\cite{amodei2016deep} and three commercial APIs (Tencent~\cite{tencent}, Alibaba~\cite{alibaba}, iFlytek~\cite{iflytek}).
For SR, we utilize the VCTK dataset~\cite{yamagishi2019cstr}. The enrollment set comprises five utterances from 81 speakers (including the target) to compute centroid embeddings, while the evaluation set consists of 800 utterances (10 each from 80 non-target speakers). We employ ECAPA-TDNN~\cite{desplanques2020ecapa}, WavLM~\cite{chen2022wavlm}, and ResNet34~\cite{he2016deep} as surrogates, and assess transferability on unseen X-vector~\cite{snyder2018x}, i-vector~\cite{dehak2010front}, and HuBERT~\cite{hsu2021hubert} models.

\noindent\textbf{Competitors.} 
We compare DUAP with four representative state-of-the-art universal adversarial attacks. For ASR, we select the targeted universal attack by Zong \etal~\cite{zong2021targeted} and the zero-query AdvDDoS framework~\cite{ge2023advddos}. For SR, we employ the embedding-disrupting method by Hanina \etal~\cite{hanina2024universal} and the robust targeted universal perturbation by Xie \etal~\cite{xie2020real}.

\noindent\textbf{Metrics.}
We use the following four metrics for evaluation. For ASR, the \textbf{Success Rate of Attack (SRoA-ASR)} is defined as the proportion of adversarial examples whose character error rate relative to the original transcription is at least $0.5$. For SR, \textbf{SRoA-SR} measures the proportion of adversarial examples that are classified as the target speaker under the CSI setting. Perceptual fidelity is assessed using the \textbf{Signal-to-Noise Ratio (SNR)}, which quantifies perturbation energy relative to the clean signal, and the \textbf{Mean Opinion Score (MOS)}, estimated via the NISQA model~\cite{mittag2021nisqa}, ranging from 1 to 5 (higher indicates better quality).

\noindent\textbf{DUAP Settings.} 
The optimization is performed via Adam with a learning rate of 0.001. Based on the ablation study in Sec.~\ref{sec:ablation}, we set the balancing weights to $\lambda_{1} = 5$ and $\lambda_{2}$ = 0.003.

\begin{table}[t]
\caption{Comparison of attack effectiveness with single-task and cross-task baselines. The lower part of each subtable lists same-task methods, while the upper part lists cross-task methods to demonstrate generalization.}
\label{tab:main_attack}
\centering

\begin{subtable}[t]{\linewidth}
\caption{Comparison with ASR methods.}
\label{tab:asr_main}
\centering
\resizebox{\linewidth}{!}{% 
\begin{tabular}{c|ccccc}
\hline
\multirow{2}{*}{\textbf{Method}} &
\multicolumn{5}{c}{\textbf{SRoA-ASR}} \\
\cline{2-6}
& Whisper & DeepSpeech2 & Tencent & Alibaba & iFlytek \\
\hline
Xie \etal~\cite{xie2020real}
& 0.007 & 0.060 & 0.000 & 0.000 & 0.073 \\
Hanina \etal~\cite{hanina2024universal}
& 0.000 & 0.013 & 0.000 & 0.000 & 0.000 \\
\hline
Zong \etal~\cite{zong2021targeted}
& 1.000 & 1.000 & \textbf{1.000} & \textbf{1.000} & \textbf{1.000} \\
AdvDDoS~\cite{ge2023advddos}
& 0.920 & 1.000 & 0.180 & 0.544 & 0.373 \\
\textbf{DUAP (ours)}
& \textbf{1.000} & \textbf{1.000} & 0.840 & 0.842 & 0.600 \\
\hline
\end{tabular}%
}
\end{subtable}

\vspace{0.8em}

\begin{subtable}[t]{\linewidth}
\caption{Comparison with SR methods.}
\label{tab:sr_main}
\centering
\resizebox{\linewidth}{!}{% 
\begin{tabular}{c|cccccc}
\hline
\multirow{2}{*}{\textbf{Method}} &
\multicolumn{6}{c}{\textbf{SRoA-SR}} \\
\cline{2-7}
& ECAPA-TDNN & WavLM & ResNet34 & HuBERT & X-vector & i-vector \\
\hline
Zong \etal~\cite{zong2021targeted}
& 0.000 & 0.000 & 0.000 & 0.000 & 0.000 & 0.000 \\
AdvDDoS~\cite{ge2023advddos}
& 0.001 & 0.034 & 0.001 & 0.001 & 0.000 & 0.586 \\
\hline
Xie \etal~\cite{xie2020real}
& 0.734 & 0.036 & 0.028 & 0.000 & 0.013 & 0.329 \\
Hanina \etal~\cite{hanina2024universal}
& 0.990 & 0.048 & 0.213 & 0.043 & 0.165 & 0.309 \\
\textbf{DUAP (ours)}
& \textbf{1.000} & \textbf{1.000} & \textbf{1.000} & \textbf{1.000} & \textbf{1.000} & \textbf{0.999} \\
\hline
\end{tabular}%
}
\end{subtable}

\vspace{-0.5em}

\end{table}

\begin{table}[t]
\caption{Comparison of imperceptibility. DUAP achieves the best fidelity among methods capable of ASR disruption.}
\label{tab:quality}
\centering
\resizebox{0.7\linewidth}{!}{% 
\begin{tabular}{c|c|cc}
\hline
\textbf{Type} & \textbf{Method} & \textbf{SNR} & \textbf{MOS} \\
\hline
\multirow{2}{*}{ASR-specific}
& Zong \etal~\cite{zong2021targeted} & -21.83 & 0.85 \\
& AdvDDoS~\cite{ge2023advddos}       & -10.22 & 1.14 \\
\hline
\multirow{2}{*}{SR-specific}
& Xie \etal~\cite{xie2020real}       & 11.89 & 1.91 \\
& Hanina \etal~\cite{hanina2024universal} & \textbf{21.73} & \textbf{2.89} \\
\hline
\textbf{Dual-Task}
& \textbf{DUAP (ours)}               & -6.96 & 1.63 \\
\hline
\end{tabular}
}
\vspace{-1.5em}
\end{table}

\subsection{Main Results}

\noindent\textbf{Attack Effectiveness.}
Table~\ref{tab:main_attack} reports the comparison results of DUAP and other competitors. 
For the ASR task, DUAP achieves a 100\% success rate on the surrogate model (Whisper) and maintains perfect transferability to DeepSpeech2. While Zong \etal~\cite{zong2021targeted} also achieve perfect attack performance, their method comes at the cost of catastrophic audio degradation (SNR of -21.8 dB), rendering the audio completely unintelligible. In contrast, DUAP maintains high perceptual quality while exhibiting significantly stronger transferability than the zero-query baseline AdvDDoS~\cite{ge2023advddos}. Specifically, it boosts the success rate on Tencent from $18.0\%$ to $84.0\%$ and on Alibaba from $54.4\%$ to $84.2\%$, validating that our targeted optimization strategy generates more robust adversarial features that generalize better to unknown black-box models.
Regarding the SR task, the superiority of DUAP is even more pronounced. While baselines like Xie \etal~\cite{xie2020real} and Hanina \etal~\cite{hanina2024universal} achieve high success rates on their source architecture (ECAPA-TDNN), they suffer from severe overfitting, with performance collapsing to below $5\%$ on heterogeneous models such as WavLM and HuBERT. Conversely, leveraged by the DNE strategy, DUAP maintains perfect or near-perfect success rates, consistently exceeding $99.9\%$ across all six models. This robust generalization spanning TDNN, Transformer, and CNN architectures confirms that DUAP captures high-level identity features that are invariant to specific model structures.

\noindent\textbf{Cross-Task Generalization.}
To justify the necessity of our dual-task formulation, we evaluate whether single-task baselines can generalize to cross-task scenarios. As shown in Table~\ref{tab:main_attack}, single-task methods exhibit severe functional limitations when applied outside their target domain. 
Specifically, SR-specific attacks such as Hanina \etal~\cite{hanina2024universal} are functionally ineffective against ASR systems, yielding success rates of nearly $0\%$ across all tested platforms and failing to disrupt transcription. 
Conversely, ASR-specific attacks like Zong \etal~\cite{zong2021targeted} fail completely in the SR task, with success rates dropping to exactly $0\%$ across all evaluated models, rendering them unable to impersonate the target identity. 
In sharp contrast, DUAP successfully bridges this gap. It is the only framework that maintains robust performance on both fronts, confirming that simply deploying a single-task attack is insufficient for compromising the entire VCS pipeline.

\noindent\textbf{Imperceptibility.}
Table~\ref{tab:quality} further evaluates the perceptual quality of the adversarial examples. Among methods capable of effective ASR disruption, DUAP achieves the best fidelity. 
DUAP significantly outperforms ASR-specific baselines: compared to Zong \etal~\cite{zong2021targeted} ($-21.8$ dB) and AdvDDoS~\cite{ge2023advddos} ($-10.2$ dB), DUAP improves the SNR to $-6.96$ dB with a MOS of $1.632$.
Although SR-specific baselines (\eg~Hanina \etal~\cite{hanina2024universal}) exhibit higher SNR, this is primarily because attacking SR systems is relatively less challenging than disrupting robust ASR models, thus requiring smaller perturbation budgets. However, such low-energy perturbations are insufficient to disrupt ASR systems (SRoA-ASR $< 5\%$).
In contrast, DUAP allocates the necessary budget to tackle the harder ASR task but effectively hides the high-energy perturbations via psychoacoustic constraints, striking the best balance between attack capability and perceptual quality.

\subsection{Ablation Study}\label{sec:ablation}

\begin{table}[!t]
    \caption{Ablation study of different loss components.}
    \label{tab:ablation}
    \centering
    \begin{tabular}{l|cc|cc}
    \hline
    \textbf{Variant} &
    \textbf{SRoA-ASR} & \textbf{SRoA-SR} & \textbf{SNR} &
     \textbf{MOS} \\
    \hline
  
    w/o $\mathcal{L}_{\mathrm{ASR}}$& 0.236 & 0.821 &  \textbf{3.296}  & \textbf{2.399} \\
    w/o $\mathcal{L}_{\mathrm{SR}}$ & 0.500 & 0.259 & -3.016  & 1.313 \\
    
    w/o $\mathcal{L}_{\mathrm{psy}}$  & 0.860 & 1.000 & -7.037  & 1.625 \\
    DUAP    & \textbf{0.856} & \textbf{0.999} & -6.964  & 1.632 \\
    \hline
    \end{tabular}%
    \vspace{-1.0em}
\end{table}

\begin{figure}[t]
    \centering
    \includegraphics[width=0.8\columnwidth]{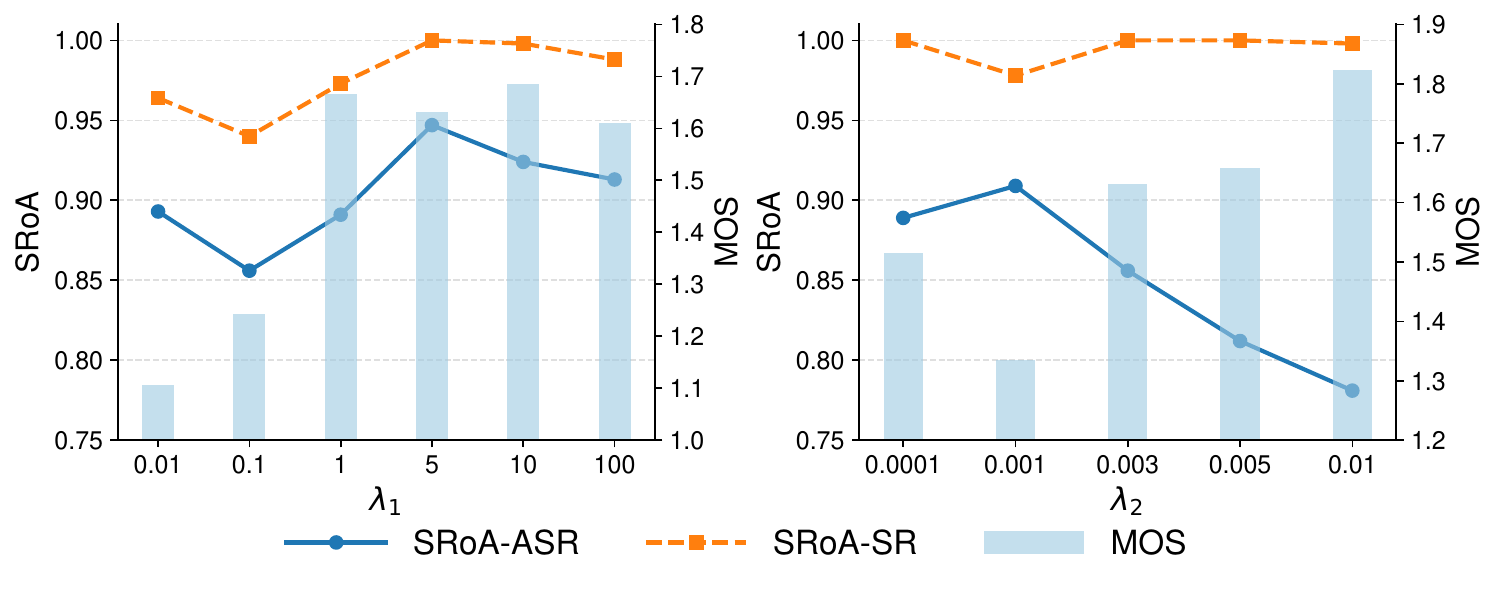}
    \caption{Impact of hyperparameters $\lambda_1$ and $\lambda_2$.
    }
    \label{fig:loss_weight_ablation}
    \vspace{-1.3em}
\end{figure}

\noindent\textbf{Variants of Our Method.}
To validate the contribution of each component, we conduct an ablation study as shown in Table~\ref{tab:ablation}. 
First, removing task-specific losses leads to a performance collapse in their respective domains. Specifically, the variant w/o $\mathcal{L}_{\mathrm{ASR}}$ yields a low SRoA-ASR of $23.6\%$, while removing $\mathcal{L}_{\mathrm{SR}}$ causes the SRoA-SR to drop to $25.9\%$. Notably, the high SNR of $3.296$ dB observed when removing $\mathcal{L}_{\mathrm{ASR}}$ confirms that disrupting robust ASR systems dominates the perturbation budget. 
Second, regarding the psychoacoustic term, the variant w/o $\mathcal{L}_{\mathrm{psy}}$ results in degraded audio quality with an SNR of $-7.037$ dB, while achieving only a negligible gain of less than $0.5\%$ in attack success. In contrast, the full DUAP framework effectively utilizes $\mathcal{L}_{\mathrm{psy}}$ to shape the perturbation, improving perceptual fidelity to a MOS of $1.632$ without compromising dual-task effectiveness.

\noindent\textbf{Hyperparameter Analysis.}
We further investigate the sensitivity of hyperparameters $\lambda_1$ and $\lambda_2$ regarding the trade-off between attack effectiveness and audio quality. As illustrated in Fig.~\ref{fig:loss_weight_ablation}, increasing $\lambda_1$ progressively enhances the SRoA-SR, which saturates near a value of 5. Consequently, we set $\lambda_1=5$ to achieve an optimal balance among SRoA-SR, SRoA-ASR, and MOS. Regarding $\lambda_2$, a larger value imposes stronger psychoacoustic constraints, resulting in higher MOS but a sharp decline in SRoA-ASR. We observe that setting $\lambda_2=0.003$ yields a favorable trade-off, maintaining robust attack performance while ensuring high perceptual fidelity.

\section{Conclusion}
In this paper, we propose DUAP, a unified adversarial attack framework capable of simultaneously compromising both ASR and SR systems. By introducing the DNE strategy and psychoacoustic constraints, DUAP effectively addresses the SR transferability challenge across heterogeneous models and achieves a superior balance between attack effectiveness and auditory imperceptibility. Extensive experiments demonstrate that DUAP successfully deceives commercial black-box APIs and various state-of-the-art speech models with a single universal perturbation, while maintaining high perceptual quality. Our work reveals critical vulnerabilities in current VCS pipelines, highlighting the urgent need for developing dual-task defense mechanisms.

\section*{ACKNOWLEDGMENT}
This research is supported in part by the Fundamental
Research Funds for the Central Universities under Grant No. 2024ZCJH05, and the Beijing Natural Science Foundation under Grant No. QY24206.
\bibliographystyle{IEEEbib}
\bibliography{arxiv}

\end{document}